%% file: output.tex
\pdfoutput=1
\documentclass[letterpaper, 10pt, conference]{ieeeconf} 
                                                         
\IEEEoverridecommandlockouts                             
\usepackage[USenglish,american]{babel}
\usepackage{amsmath}
\usepackage{amsfonts}
\usepackage[usenames,dvipsnames]{color}
\usepackage{xcolor}
\usepackage{algorithmic}
\usepackage{algorithm}
\usepackage{array}
\usepackage[caption=false,font=footnotesize]{subfig}
\usepackage{textcomp}
\usepackage{stfloats}
\usepackage{url}
\usepackage{verbatim}
\usepackage{graphicx}
\usepackage{soul}
\usepackage{cite}
\usepackage{bm}
\usepackage{tikz}
\usepackage{pgf}
\usepackage{soul}
\usepackage{comment}
\usepackage{hhline}
\usepackage{multirow}
\usepackage{siunitx}
\usepackage{hyperref}

\usepackage[utf8]{inputenc}
\usepackage{pgfplots}
\DeclareUnicodeCharacter{2212}{−}
\usepgfplotslibrary{groupplots,dateplot}
\usetikzlibrary{patterns,shapes.arrows}
\pgfplotsset{compat=newest}

\def\smpcvpm{{SMPC+CVPM}}
\def\smpcftp{{SMPC+FTP}}
\def\Rset{{\mathbb{R}}}
\def\xii{{\bm \xi}}
\def\Xii{{\bm \Xi}}
\def\uu{{\bm u}}
\def\UU{{\bm U}}
\def\Aev{{\bm A_\text{d}}}
\def\Bev{{\bm B_\text{d}}}
\def\xido{{\bm \xi^\text{DO}}}
\def\udo{{\bm u^\text{DO}}}

\newtheorem{remark}{Remark}
\newtheorem{definition}{Definition}

\input{space.tex}

\begin{document}

\title{Safe and Non-Conservative Trajectory Planning for Autonomous Driving Handling Unanticipated Behaviors of Traffic Participants}%

\author{Tommaso Benciolini*, Michael Fink*, Nehir Güzelkaya, Dirk Wollherr, Marion Leibold%
\thanks{* The first two authors contributed equally to this work.}
\thanks{This work is funded by the Deutsche Forschungsgemeinschaft (DFG, German Research Foundation) with the project number 490649198.}
\thanks{T. Benciolini,  M. Fink, N. Güzelkaya, D. Wollherr, and M. Leibold are with the Chair of Automatic Control Engineering at the Technical University of Munich, Germany (email: {\tt\small\{t.benciolini; michael.fink; nehir.guezelkaya; dw; marion.leibold\}@tum.de}).}%
}

\maketitle

\begin{abstract}
Trajectory planning for autonomous driving is challenging because the unknown future motion of traffic participants must be accounted for, yielding large uncertainty. Stochastic Model Predictive Control (SMPC)-based planners provide non-conservative planning, but do not rule out a (small) probability of collision. We propose a control scheme that yields an efficient trajectory based on SMPC when the traffic scenario allows, still avoiding that the vehicle causes collisions with traffic participants if the latter move according to the prediction assumptions. If some traffic participant does not behave as anticipated, no safety guarantee can be given. Then, our approach yields a trajectory which minimizes the probability of collision, using Constraint Violation Probability Minimization techniques. Our algorithm can also be adapted to minimize the anticipated harm caused by a collision. We provide a thorough discussion of the benefits of our novel control scheme and compare it to a previous approach through numerical simulations from the CommonRoad database.
\end{abstract}

\section{Introduction}
\vspace{-11.6cm}
\noindent\hspace*{0.1\textwidth}
\parbox[t]{20cm}{\small This~work~has~been~accepted~to~the~IEEE~2024~International~Conference~on~Intelligent~Transportation~Systems.}
\vspace{10.68cm}

An essential component enabling autonomous driving are motion planning algorithms that handle the interplay with other vehicles, pedestrians, and cyclist, referred to as Dynamic Obstacles (DO) in this work. 
Trajectory planning algorithms must account for the unknown future behavior of DOs in a safe yet non-conservative, human-like manner.

Machine learning approaches are very popular for their ability to replicate human-like motion using past data~\cite{kumar2013,phillips2017,rosolia2017}. However, giving formal guarantees of collision avoidance with machine learning methods is challenging.

Model Predictive Control (MPC) is a popular algorithm for trajectory planning~\cite{levinson2011}, which realizes a human-like control strategy~\cite{carvalho2014,gutjahr2017}, while still allowing to provide formal guarantees. 
In MPC the trajectory of the automated vehicle, here Ego Vehicle (EV), is computed by iteratively solving an Optimal Control Problem (OCP) over fixed-length receding horizons. Desired control objectives are given as a cost function and safety conditions and traffic rules, together with physical limitations of the vehicle, result in a set of constraints. At each sampling time the action optimal sequence of the EV for the full prediction horizon is obtained. The first input of the optimal sequence is applied and the process is repeated when new measurements of the environment are collected, adapting the future trajectory to the changing traffic configuration.

The unknown future trajectory of DOs is a source of uncertainty that must be accounted for. The uncertain future motion of DOs can be described in terms of worst-case movements underlying a given intention and considering the traffic rules. Then, the EV trajectory is \textit{safe} if it is collision free for all realizations of the uncertain DO future motion within the worst-case assumptions. Furthermore, a probability distribution for movements within the worst-case outcomes can be assumed. The key challenge from a control perspective lies in balancing the need for safety and avoiding overly conservative behaviors typical of robust approaches based on a \textit{robust} worst-case analysis~\cite{dixit2020,soloperto2019,sontges2018,gillula2013,schurmann2018}. Furthermore, robust approaches are only applicable if the assumed uncertainty is bounded and small and come at the price of a significant conservatism in the trajectory planning of the EV, leading to considerable inefficiency in environments with large uncertainty about the future motion of DOs. To reduce conservatism, Stochastic Model Predictive Control (SMPC)~\cite{mesbah2016,farina2016} has been proposed, reformulating collision avoidance constraints in chance (probabilistic) constraints, that must only be satisfied up to a user-specified probability level. SMPC yields a non-conservative EV trajectory~\cite{carvalho2014}, but produces open-loop predictions with non-zero probability of collision that may consequently result in collisions in closed loop. Therefore, SMPC does not provide safety guarantees typical of robust approaches. A scheme to take advantage of the optimistic planning of SMPC while still giving collision avoidance guarantees has been proposed in our previous work~\cite{brudigam2023}. However, the scheme requires the cascaded solution of two optimal control problems at each iterations, which is impractical in real-world applications with small sampling times. Furthermore, the approach relies on assumptions for the worst-case behavior of DOs, without addressing the case in which such assumptions are violated. Differently, contingency MPC~\cite{alsterda2019} has been proposed, in which non-conservative yet safe planning is achieved by computing both a nominal and a safe input sequence that produce the same first-step prediction. Although the cascaded solution of two cascaded optimization problems, computing the two input sequences at once does significantly increase the dimension of the optimization problem. Furthermore, the algorithm does not allow to use the full SMPC trajectory even when it is safe to do so, since the first element of the input sequence is always optimized jointly with the robust trajectory.

In this work we propose a novel control scheme for efficient and safe trajectory planning for automated driving that leverages on the optimistic SMPC planning still yielding a collision-free motion. At first, the SMPC OCP is solved, yielding an optimistic non-conservative prediction for the EV trajectory. Then, the first predicted state of the SMPC trajectory is used as initial condition for a further planning problem, \textit{robust} with respect to the uncertain DO motion. The solution to this problem is not computed, the problem only serves as certification of safety. If such further OCP admits a solution, the first SMPC input is guaranteed not to lead to unavoidable collisions. Therefore, in this case, the SMPC input is considered safe and is applied. 
By contrast, if a solution to the further robust control problem does not exist, i.e., the robust control problem is infeasible, the SMPC solution is disregarded, since it could lead to collisions that can not be prevented in the next time steps. If the SMPC solution fails the safety certification, we plan a robust trajectory from the current initial condition, which exists because of the safety certification of the previous iteration. Thus, the EV does not cause collisions with DOs.

In practice, the actual motion of DOs might violate the worst-case assumptions. We refer to these DO motions as \textit{unanticipated} behaviors. Also illegal behaviors, where a DO violates the traffic rules, are considered unanticipated behaviors. In presence of unanticipated behaviors of DOs the safety guarantee from combining SMPC with a robust backup does not hold. In fact, there might be time instants in which the SMPC solution is not certified to be safe, that is, the robust problem from the first predicted step is infeasible, and furthermore also the robust planner for the current initial condition could fail, that is, it could not be possible to plan a trajectory that is collision-free. In such situations, we plan the EV trajectory by minimizing the probability of collision over the prediction horizon. In our approach, we implement the robust planner and the probability-minimization planner using the Constraint Violation Probability Minimization (CVPM) approach~\cite{brudigam2021c,fink2022}: when feasible, a control action yielding zero probability of collision is applied (\textit{robust case}); otherwise, the probability of collision is minimized (\textit{probabilistic case}).

Our novel scheme, named \smpcvpm, is computationally efficient since it is not necessary to compute the robust backup after the solution of the SMPC problem. Instead, the solution of the cascaded OCP is replaced by a fast set computation to check the existence of the robust solution, and the logical structure can benefit from parallelization. Furthermore, addressing those situations in which a DO does not behave as anticipated greatly enhance the applicability of the algorithm. In fact, most other planners do not address these situations and just apply previously stored backup plans~\cite{brudigam2023,alsterda2019}; however, in presence of unanticipated behaviors the stored backup is not supported by safety guarantees, nor it is designed to react to the unforeseen situation according to a reasonable criterion. We validate the algorithm and discuss the properties through simulations on benchmark scenarios from the CommonRoad dataset~\cite{althoff2017}. The simulation code is available at \href{https://github.com/tbenciolini/Safe-and-Non-Conservative-Trajectory-Planning}{https://github.com/tbenciolini/Safe-and-Non-Conservative-Trajectory-Planning}.

The remainder of the paper is organized as follows. Sec.~\ref{sec:preliminaries} recalls the dynamical models used for the EV and the dynamic obstacles and presents the MPC-based trajectory planners that are considered in the control scheme. Sec.~\ref{sec:smpcvpm} introduces our novel \smpcvpm~scheme, discussing its properties and giving an overview of the implementation details. An efficient way of solving the probability minimization is given in Sec.~\ref{sec:prob_minimization_implementation}. Sec.~\ref{sec:simulations} presents the simulations from the CommonRoad database. Sec.~\ref{sec:discussion} and~\ref{sec:conclusion}, respectively, offer a discussion and conclusive remarks.

\section{Preliminaries}
\label{sec:preliminaries}
In the following we outline the dynamical models used for the EV and for DOs, Sec.~\ref{sec:dynamic_models}. Then, we outline the SMPC-based trajectory planners (Sec.~\ref{sec:mpc_based_planning}) and CVPM (Sec.~\ref{sec:cvpm}). 

\subsection{Models of Ego Vehicle and Dynamic Obstacles}
\label{sec:dynamic_models}
In this section we introduce the dynamical models used to make predictions of the EV trajectories and of DOs. For the EV, we use a curvature-aware kinematic bicycle model referred to the road-aligned Frenet coordinate system~\cite{werling2008}. The state of the EV $\xii=\left[s,d,\phi,v\right]^\top$ consists of the longitudinal and lateral positions of the center of mass of the EV in the road-aligned reference frame, $s$ and $d$, respectively, of the yaw angle $\phi$, and of the linear velocity $v$. The input of the EV $\uu=\left[a,\delta\right]^\top$ consists of the acceleration $a$ and the front steering angle $\delta$. In the following, an approximated discrete-time model is~used
\begin{equation}
    \xii_{k+1}=\Aev(\xii_k-\xii_0)+\Bev\uu_k+\bm c_\text{d},
    \label{eqn:linearized_discretized_evmodel}
\end{equation}
that is obtained linearizing the kinematic bicycle model around the current state $\xii_0$ and zero input $\uu^\ast=\left[0,0\right]^\top$, and discretizing with zero-order hold of sampling time $T$. All details about the linearization and discretization procedure, together with the explicit analytical expression of ${\Aev\in\Rset^{4\times4}},\ {\Bev\in\Rset^{4\times2}}$, and $\bm c_\text{d}\in\Rset^4$ from~\eqref{eqn:linearized_discretized_evmodel}, are given in our previous work~\cite{benciolini2023}. Such model couples the longitudinal and lateral dynamics of the vehicle and allows efficient computation. However, it does not accurately represent some of the dynamics features, such as the nonlinear tire characteristics, and the linearization introduces further approximations, that we neglect because within this work, we primarily focus on the uncertainty introduced by the unknown future trajectory of DOs. Different, more accurate models can also be used without compromising the validity of the presented approach. However, the considerations on the computational complexity given in the following hold true only for linear models. Similarly, uncertainty about the measurements of the EV state might be considered and accounted for in the derivation of the safety collision-avoidance constraints without significant changes in the algorithm.

For the assumed dynamics of the DOs, we adopt a double integrator model with decoupled dynamics for the longitudinal and lateral components. For ease of notation, in the following we refer to one DO only. However, all considerations hold similarly for multiple DOs close enough to the EV. The state of each DO is ${\xido = \left[s^\text{DO},v_s^\text{DO},d^\text{DO},v_d^\text{DO}\right]^\top}$, consisting of longitudinal and lateral positions and velocities, and the input ${\udo = \left[a_s^\text{DO}, a_d^\text{DO}\right]^\top}$ comprises the longitudinal and lateral acceleration, $a_s^\text{DO}$ and $a_d^\text{DO}$, respectively. We model the action of DO drivers as an LQR-based controller tracking a constant reference state that expresses the intention of the driver, for example maintaining a given cruise speed or a lateral target position. The DO prediction model is thus denoted as
\begin{subequations}
\begin{align}
\bm\xi^\text{DO}_{k+1} &= \bm A^\text{DO}\bm\xi^\text{DO}_k+\bm B^\text{DO}\bm u^\text{DO}_k\\
\bm u^\text{DO}_k &= \bm K^\text{DO}\left(\bm\xi^\text{DO}_k-\bm\xi^\text{DO}_{\text{ref},k}\right)+\bm w^\text{DO}_k,
\end{align}
\label{eqn:DOmodel}%
\end{subequations}
where $\bm A^\text{DO}$ and $\bm B^\text{DO}$ are the discrete-time point-mass model matrices as in~\cite{brudigam2023,benciolini2023}. The input, which is bounded
\begin{equation}
\bm u^\text{DO}_\text{min}\leq\udo\leq\bm u^\text{DO}_\text{max},
\label{eqn:DOlimits}
\end{equation}
consists of a deterministic feedback loop and of a stochastic term. On the one hand, the deterministic term is designed to steer the state towards the reference state $\bm\xi^\text{DO}_{\text{ref},k}$, with a rate depending on the LQR-based controller gain $\bm K^\text{DO}$. This term represents the likely nominal trajectory being pursued. On the other hand, we account for uncertainty about the input for such nominal trajectory. Within the scope of this work, the uncertainty models the unpredictability of the human decision and the fact that the human intention is not known. Therefore, we consider the disturbance $\bm w^\text{DO}$ as part of the input. Such uncertainty is modeled as truncated Gaussian, $\bm w^\text{DO}\sim\mathcal{N}\left(0,\bm\Sigma^{\bm w,\text{DO}},\mathcal{W}^\text{DO}\right)$. Moreover, we assume that measurements of the DO states available to the EV are corrupted by the measurement noise $\bm v^\text{DO}$, which is a zero-mean truncated Gaussian distribution $\bm v^\text{DO}\sim\mathcal{N}\left(0,\bm\Sigma^{v,\text{DO}},\mathcal{V}^\text{DO}\right)$.
In this work, stochastic and robust control approaches are combined, therefore the assumption of bounded support for the distributions is later relaxed and unbounded support is assumed for the stochastic control approaches.

\begin{definition}
\label{def:unanticipated_behavior}
An unanticipated behavior of DOs is any trajectory that is not in the reachable space of model~\eqref{eqn:DOmodel} subject to constraints~\eqref{eqn:DOlimits}, that is, trajectories that cannot be obtained from dynamics~\eqref{eqn:DOmodel} for any input satisfying~\eqref{eqn:DOlimits}.
\end{definition}

\begin{remark}
In practice, the actual DO control policy and intended behavior and are not known to the EV and therefore the assumptions could be violated. Thus, the bounds on the DO input~\eqref{eqn:DOlimits} and the Gaussian probability distribution for the stochastic component of the DO input are to be intended as the assumption of the EV on the DO model, on what is considered an anticipated behavior.
\end{remark}

Such unanticipated behaviors represent both, future trajectories that are not feasible given the assumed physical limitations of the DOs and future trajectories that violate traffic rules. These assumed models are used to derive safe areas for the EV in the MPC schemes presented in the following sections. More sophisticated prediction models that more precisely characterize unanticipated behaviors, possibly taking traffic rules into account, could be used without major modification of the proposed approach. For our control architecture, which is the main contribution of this work, it is sufficient that prediction models result in a bounded set of possible future positions of DOs.

\subsection{Stochastic Model Predictive Control}
\label{sec:mpc_based_planning}
In trajectory planning for automated driving, constraints account for the physical limitations of the EV, for the traffic rules, and for collision avoidance safety constraints.
However, the future positions of DOs are not known deterministically, rather, nominal predictions are affected by uncertainty. 
The introduced schemes consider the same prediction horizon length $N$ and share the same cost function
\begin{equation}
J(\bm\xi_0,\UU_N) = \|\Delta\xii_{N}\|^2_{\bm{P}}+\sum_{k=0}^{N-1}\left(\|\Delta\xii_k\|^2_{\bm{Q}}+\|\bm{u}_k\|^2_{\bm{R}}\right).
\label{eqn:costfcn}
\end{equation}
The cost $J$ is designed to penalize large inputs, where $\bm{R}\succ0$. 
Large deviations ${\Delta\xii=\xii-\xii^\ast}$ of the EV from the desired reference $\xii^\ast$ are penalized, with ${\bm{Q},\bm{P}\succeq0}$.

The schemes presented in the following differ in the formulation of safety constraints. In this work, for simplicity we refer to $\bm x$ as the current state of the environment, considering the current states of the EV and DOs, i.e. $ \bm x = \left[\xii^\top,{\bm\xi^\text{DO}}^\top\right]^\top$.
Therefore, we refer with $\text{SMPC}(\bm x)$ and $\text{CVPM}(\bm x)$ to problems
evaluated for the same current traffic configurations, i.e., only differing in the formulation of the safety constraints.

In the SMPC approach, hard constraints for collision avoidance are relaxed and must be satisfied only up to a user-specified level of probability, $\beta<1$. The OCP considered at each sampling time is
\begin{subequations}
\begin{alignat}{2}
    \text{SMPC}(\bm x):\ \min_{\UU_N}&\ J(\bm\xi_0,\UU_N)\label{eqn:smpc_cost_function}\\
    \text{s.t. }\xii_{k+1}&=\bm f(\xii_k,\uu_k),\ &&\forall k=0,\dots,N-1\label{eqn:smpc_model}\\
    \xii_k&\in\mathcal{X}_k,\ &&\forall k=1,\dots,N\label{eqn:smpc_state_constraints}\\
    \uu_k&\in\mathcal{U}_k,\ &&\forall k=0,\dots,N-1\label{eqn:smpc_input_constraints}\\
    \Pr[\xii_k&\in\mathcal{S}(\bm \xi^\text{DO}_k)]\geq \beta,\ &&\forall k=1,\dots,N,\label{eqn:smpc_safety_constraints}
\end{alignat}%
\label{prb:smpc}%
\end{subequations}%
where $N$ is the prediction horizon, ${\UU_N=\left[\uu_0^\top,..,\uu_{N-1}^\top\right]^\top}$, $\bm f$ in~\eqref{eqn:smpc_model} is a compact representation of~\eqref{eqn:linearized_discretized_evmodel}, and $\mathcal{X}_k$ and $\mathcal{U}_k$ in~\eqref{eqn:smpc_state_constraints} and~\eqref{eqn:smpc_input_constraints}, respectively, are state and input hard constraints due to the physical limitations of the EV. %

Constraints~\eqref{eqn:smpc_safety_constraints} are probabilistic chance constraints: the EV must meet the safety requirements $\mathcal{S}(\bm \xi^\text{DO}_k)$ with respect to the future positions of the DOs. The set of safe positions $\mathcal{S}(\bm \xi^\text{DO}_k)$ is obtained considering the probability distribution of the uncertainties $\bm w^\text{DO}$, i.e., the unknown future behavior of DOs, and $\bm v^\text{DO}$, i.e., the measurement noise. To handle chance constraints numerically, we utilize unbounded disturbances and  reformulate~\eqref{eqn:smpc_safety_constraints} analytically using the assumptions on the probability distribution of the uncertainty.  
Observe that the SMPC approach allows a $1-\beta$ probability of violation in open loop for each time step $k$ in the prediction horizon, that is, in the predictions a collision is possible.%


\subsection{Constraint Violation Probability Minimization}
\label{sec:cvpm}
In the following, we recall the concept of CVPM which was introduced in~\cite{brudigam2021c} and~\cite{fink2022}. 
CVPM switches between two optimization problems to enable a robust solution as long as possible by solving the problem $\text{CVPM Robust}(\bm x)$, denoted as \textit{robust case}. 
If $\text{CVPM Robust}(\bm x)$ is not feasible, the \textit{probabilistic case} is applied, where $\text{CVPM Prob}(\bm x)$ is solved and minimizes the probability of constraint violation. 

CVPM considers the set of safe EV state sequences as a Cartesian product of safe position sets, i.e., ${\hat{\mathcal{S}}\left(\bm \Xi^\text{DO}_N\right) = \prod_{k=1}^N\mathcal{S}(\bm \xi^\text{DO}_k)}$, for a given DO state sequence $\bm\Xi_N^\text{DO}= [{\xii_1^\text{DO}}^\top,\dots,{\xii_N^\text{DO}}^\top]^\top$.  The set  $\hat{\mathcal{S}}\left(\bm \Xi^\text{DO}_N\right)$ is implemented as in~\cite[Sec.\;V-B]{brudigam2023}.
The robust case OCP is
\begin{subequations}
\begin{alignat}{2}
\text{CVPM Robust}&(\bm x):\ \min_{\UU_N}\ J(\bm\xi_0,\UU_N)\label{eqn:cvpm_cost_fcn}\\
    \text{s.t. }\xii_{k+1}&=\bm f(\xii_k,\uu_k),\   &&\hspace{-5mm}\forall k=0,\dots,N-1\\
    \xii_k&\in\mathcal{X}_k,\                       &&\hspace{-5mm}\forall k=1,\dots,N\\
    \uu_k&\in\mathcal{U}_k,\                        &&\hspace{-5mm}\forall k=0,\dots,N-1\\
    &\hspace{-5mm}\Pr\Big[\Xii_k\notin\hat{\mathcal{S}}(\hat{\bm\Xi}^\text{DO}_k)\Big]=0 . \label{eqn:cvpm_safe_constraint}
\end{alignat}%
\label{prb:smpc_safe}%
\end{subequations}%
Constraint \eqref{eqn:cvpm_safe_constraint} imposes that the EV state sequence satisfies the collision avoidance constraint robustly.

In the  \textit{probabilistic case}  a solution with zero probability of constraint violation does not exist. Therefore, as in~\cite{brudigam2021c}, we minimize the probability of constraint violation over the whole prediction horizon by solving the following OCP
\begin{subequations}
\begin{alignat}{2}
\text{CVPM Prob}&(\bm x):\ \min_{\UU_N}\ \Pr\big[\Xii_k&&\notin\hat{\mathcal{S}}(\hat{\bm\Xi}^\text{DO}_k)\big]\label{eqn:cvpm_probablistic_case}\\
    \text{s.t. }\xii_{k+1}&=f(\xii_k,\uu_k),\ &&\forall k=0,\dots,N-1\\
    \xii_k&\in\mathcal{X}_k,\ &&\forall k=1,\dots,N\\
    \uu_k&\in\mathcal{U}_k,\ &&\forall k=0,\dots,N-1.
\end{alignat}%
\label{prob:cvpm_probablistic_case}%
\end{subequations}%
More details on the minimization of the probability of constraint violation in \eqref{eqn:cvpm_probablistic_case} are given in Sec.~\ref{sec:prob_minimization_implementation}.

The distinction into two different cases, introduced in~\cite{fink2022}, is outlined in the following. 
The \textit{robust case} is applied if 
\begin{equation}
    \exists \UU_N: \Pr\left[\Xii_N\notin\hat{\mathcal{S}}\left(\bm \Xi^\text{DO}_N\right)\right]   = 0.
\end{equation}
To evaluate this condition in practice, the set of feasible input trajectories for the robust case
\begin{IEEEeqnarray}{c}
    \label{eqn:u_robust}
     \mathcal{U}_\text{robust} \!=\! \left\lbrace\! \UU_N \middle|\Xii_N\!\in\!\hat{\mathcal{S}}\!\!\left(\bm \Xi^\text{DO}_N\right) \!,\! 
     \forall  \bm w^\text{DO}_k  \!\in\!  \mathcal{W}^\text{DO},
     \forall \bm v^\text{DO}_k  \!\in\! \mathcal{V}^\text{DO}
     \!\right\rbrace
     \IEEEeqnarraynumspace
\end{IEEEeqnarray}
is determined. 
A fast linear program with $\mathcal{U}_\text{robust}$ as a constraint is used to determine whether it is empty \cite{borrelli2017}. If $ \mathcal{U}_\text{robust}$ is nonempty, the linear program is feasible, and the robust case is applied. If $ \mathcal{U}_\text{robust}=\emptyset$, the probabilistic case is used. 
The solutions to the problems $\text{CVPM Robust}(\bm x)$ and $\text{CVPM Prob}(\bm x)$  yield the optimal input sequence~$\UU_N^*$, where the first element is applied to the system. 

\begin{remark}
In SMPC, chance constraints \eqref{eqn:smpc_safety_constraints} are defined separately for each time step, to allow constraint tightening.
In CVPM, the joint probability distribution is used for minimization, because reducing too much the probability of constraint violation in the first steps might increase the probability of constraint violation for further steps. 
\end{remark}

\section{Novel \smpcvpm~Control scheme}
\label{sec:smpcvpm}
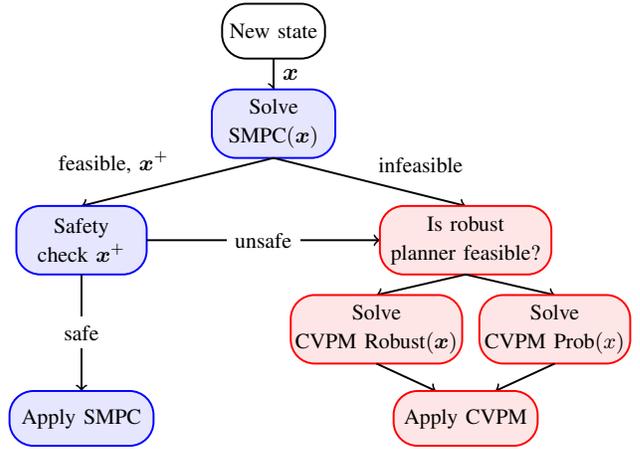
\begin{figure}
    \centering
    \resizebox{\columnwidth}{!}{
    \input{Figs/Diagram_logic} 
    }
    \caption{Diagram of the logic of the novel \smpcvpm~scheme.}
    \label{fig:smpcvpm_diagram_logic}
\end{figure}

In this section we outline the proposed novel \smpcvpm~scheme and discuss its properties. SMPC plans a non-conservative EV trajectory. However, it accepts a non-zero risk of collision. We use CVPM to evaluate the safety of the SMPC trajectory. Furthermore, CVPM is also used to plan an alternative trajectory, that is safe for any realization of the uncertainty. If the real behavior of DOs violates the assumptions of the EV, then CVPM returns a trajectory which minimizes the probability of a collision.

The logic to decide whether  SMPC or CVPM must be applied is illustrated in Figure~\ref{fig:smpcvpm_diagram_logic}. At each time step, at first, the $\text{SMPC}(\bm x)$ OCP is solved, that is, an optimistic trajectory based on the current traffic configuration~$\bm x$ is computed. If a solution to $\text{SMPC}(\bm x)$ exists, the predicted next step is considered, that is, the traffic configuration at the next sampling time, $\bm x^+$ resulting from the first input of the $\text{SMPC}(\bm x)$ solution. From $\bm x^+$, a safety check is performed considering the prediction horizon from prediction step $k=1$ to $k=N+1$. The goal of the safety check is to verify whether after applying the solution of $\text{SMPC}(\bm x)$, at the next time step it is still possible to determine a safe trajectory accounting for the worst case movement of the DOs, following a robust approach. Precisely, the existence of a feasible solution to OCP $\text{CVPM Robust}(\bm x^+)$~\eqref{prb:smpc_safe} is evaluated. Yet, a solution to $\text{CVPM Robust}(\bm x^+)$ must not be determined at present, in fact OCP~\eqref{prb:smpc_safe} is not solved; rather, only the existence of a solution is discussed by performing a set computation and therefore the procedure is computationally undemanding. If the safety check succeeds, SMPC will lead to a safe state from which a feasible robust solution can be obtained at the next sampling time. In such case, the first input of the solution to $\text{SMPC}(\bm x)$ is applied.

\begin{definition}
The solution to the $\text{SMPC}(\bm x)$ OCP is safe if the safety check for $\bm x^+$ succeeds, that is, if it is certified that applying the SMPC solution leads to a state $\bm x^+$ from which a collision free trajectory can be planned, if DOs behave consistently with the assumptions in Sec.~\ref{sec:dynamic_models}.
\end{definition}

\begin{remark}
The safety check in particular verifies the safety of the initial condition of the following OCP. Therefore, if the first predicted state of the SMPC trajectory does not satisfy collision avoidance constraints in a robust sense, the safety check fails as well. Rigorously speaking, the safety check succeeds if and only if $ \mathcal{U}_\text{robust} \neq \emptyset$ for $\bm x^+$, with $ \mathcal{U}_\text{robust}$ from~\eqref{eqn:u_robust}, but also including initial condition in the state sequence $\Xii_N$.
\end{remark}

\begin{remark}
The predicted traffic configuration $\bm x^+$ used for the safety check is not the prediction used in the SMPC scheme, based on the approximated EV model~\eqref{eqn:linearized_discretized_evmodel}. Rather, a nonlinear and continuous-time model is used to compute a more accurate prediction of the future state of the EV if the SMPC input is applied. Since the safety check is not part of the SMPC OCP, the predicted traffic configuration $\bm x^+$ is computed only once per iteration, after that the SMPC OCP has been solved, and a continuous-time nonlinear model does not severely affect the computation time of the scheme. Yet, the prediction models of DOs must be consisted with the assumptions in Sec.~\ref{sec:dynamic_models}.
\end{remark}

Since the $\text{SMPC}(\bm x)$ is not recursively feasible, sometimes it does not yield a solution. Furthermore, even when a solution exists, this might not be safe, that is, applying the first input prevents the existence of a solution to the robust planner for the first predicted step. In either of these situations, the OCP $\text{CVPM}(\bm x)$ is solved, as shown in the right branch in Figure~\ref{fig:smpcvpm_diagram_logic}. First, we check which case of the CVPM algorithm must be used. If possible, the robust case is solved, producing a trajectory that is collision-free for any realization of the uncertainty within the assumptions from Sec.~\ref{sec:dynamic_models}. If this is not possible, the probabilistic case is used to obtain a trajectory which minimizes the probability of collision.

\begin{remark}
If $\text{SMPC}(\bm x)$ does not yield a solution, it is unlikely that the robust case of $\text{CVPM}(\bm x)$ is applicable, since the latter is a robust planner with tighter constraints. Still, because of minor differences in the derivation of the safety constraints, this cannot be completely ruled out for rare cases. Nevertheless, the robust case of $\text{CVPM}(\bm x)$ is mainly used to plan a safe trajectory when the solution to $\text{SMPC}(\bm x)$ exists but is not safe, i.e., if it fails the safety check.
\end{remark}

The \smpcvpm~scheme allows to benefit from the efficient and optimistic SMPC planning when it is safe to do so, i.e., if a safe trajectory can be obtained for the predicted traffic configuration $\bm x^+$. Yet, the CVPM robust case computes a safe trajectory if the SMPC solution is not safe. Thus, if DOs move as expected according to the assumed models (Sec.~\ref{sec:dynamic_models}), our scheme yields always a collision-free trajectory.

Moreover, the proposed scheme addresses situations where the EV must deal with unforeseen behaviors of DOs, whose motion violates the hypotheses on which the robust motion planner is built. This might happen, for example, due to unanticipated behaviors of traffic participants (Definition~\ref{def:unanticipated_behavior}). If the actual motion of the DOs violates the worst-case hypotheses on the disturbance distributions considered in the robust approach, the previously established safety guarantees do not longer hold. In such cases, the EV re-plans a new safe trajectory if possible; if not possible, the EV aims at minimizing the probability of a collision due to the possibly unlawful behavior of other traffic participants. Observe that if DOs behave as expected, the EV returns a collision-free trajectory, therefore possible collisions are due to the DO unanticipated behaviors, not caused by the EV.

The logical structure of the novel~\smpcvpm~scheme is computationally efficient and can be parallelized. In fact, the CVPM branch (in red in Figure~\ref{fig:smpcvpm_diagram_logic}) is independent of the solution of the SMPC branch, therefore can be computed in parallel and only applied if necessary. Furthermore, the safety check for $\bm x^+$, which follows the SMPC OCP, only consists of a set computation, since no safety backup must be computed or stored, only the existence of a solution must be evaluated. As a result, the solution of two cascaded optimization problems is never required to obtain the solution.

\section{Probability Minimization}
\label{sec:prob_minimization_implementation}
In the \textit{probabilistic case} of CVPM, the goal is to minimize the constraint violation probability as denoted in \eqref{prob:cvpm_probablistic_case}. 
In this section, we reformulate the minimization to solve a slightly different but efficient problem. 
Optimizing the probability in~\eqref{prob:cvpm_probablistic_case} requires numerical integration over a multivariate Gaussian, which is tedious and limits online usability. Hence, the problem is approximated with a quadratic program. 

We first specify the set of feasible state sequence $\hat{\mathcal{S}}\left(\bm \Xi^\text{DO}_N\right)$, derived by the  procedure from~\cite{brudigam2023}, in a half space representation with $n_\text{q}$ inequalities, i.e., 
\begin{align}\label{eqn:feasible_set}
     \hat{\mathcal{S}}\left(\bm \Xi^\text{DO}_N\right)
    =\left\lbrace \Xii_N  \;\middle|\;\bm{Q}_N\Xii_N+\bm{q}_N \leq 0 \right\rbrace,
\end{align}
where $\bm{Q}_N \in \mathbb{R}^{n_\text{q}\times 4N}$, $\bm{q}_N \in \mathbb{R}^{n_\text{q}}$ are the inequality matrix and the offset vector, respectively, and both are functions of the predicted DO state sequence.
Since the DO state sequence $\bm \Xi^\text{DO}_N$ is affected by the disturbance $\bm w^\text{DO}_k$ and noise $\bm v^\text{DO}_k$, the set $\hat{\mathcal{S}}\left(\bm \Xi^\text{DO}_N\right)$ is uncertain, which is modeled with a random offset vector   $\bm{q}_N$ following the normal distribution $\bm{q}_N\sim \mathcal{N}(\bar{\bm{q}}_N , \Sigma^{\bm{q}_N})$ with the mean $\bar{\bm{q}}_N $ and covariance~$\Sigma^{\bm{q}_N}$.
The optimization of \eqref{eqn:cvpm_probablistic_case} is efficiently solved with
\begin{subequations}
\begin{alignat}{2}
    \text{CVPM Prob}&(\bm x):\ \min_{\UU_N,  \bm S_N} \| \bm S_N&&-\bm{\mu}_N \|^2_{\left(\bm\Sigma^{\bm{q}_N}\right)^{-1}}\label{eqn:approx_probabilistic_case}\\
    \text{s.t. }\xii_{k+1}&=f(\xii_k,\uu_k),\ &&\forall k=0,\dots,N-1\\
    \xii_k&\in\mathcal{X}_k,\ &&\forall k=1,\dots,N\\
    \uu_k&\in\mathcal{U}_k,\ &&\forall k=0,\dots,N-1\\
    \bm{\mu}_N &= \bm{Q}\Xii_N +  \bar{\bm{q}}_N,\\
    \bm{S}_N& \leq 0.
\end{alignat}%
\label{prb:approx_cvpm_prob_case}%
\end{subequations}%
The decision variable $\bm{S}_N\in\mathbb{R}^{4N}$ in~\eqref{eqn:approx_probabilistic_case} is a non-positive slack variable. Problem \eqref{prb:approx_cvpm_prob_case} is a quadratic problem that is solved efficiently, yielding a solution of the optimal input sequence~$\UU_N^*$ to the original problem \eqref{prob:cvpm_probablistic_case}.

In the following, we derive the approximated quadratic program \eqref{prb:approx_cvpm_prob_case}, by deriving the probabilistic distribution of the offset $\bm{q}_N$. 
We start by decomposing the state sequence of the DO~$\bm \Xi^\text{DO}_N$  in a nominal component~$\bar{\bm\Xii}^\text{DO}_k$ and the stochastic component based on the disturbances~$\bm w^\text{DO}_k$. 
For the minimization of the constraint violation probability, the untruncated Gaussian distribution of the disturbance is used, along the lines of~\cite{brudigam2021c,fink2024}.  
The nominal component $\bar{\bm\Xii}^\text{DO}_N $ is determined with a prediction based on~\eqref{eqn:DOmodel}
resulting in the random prediction 
\begin{align}
   \bm\Xii^\text{DO}_N   &= \bar{\bm\Xii}^\text{DO}_N  + \bm B^\text{DO}\bm W^\text{DO}_{N},
\end{align} 
where $\bm B^\text{DO}$ is defined as in~\cite[Eq. (8.5)]{borrelli2017}. 
The covariance matrix of the disturbance sequence given as ${\bm \Sigma^{\bm W,\text{DO}}  =  \operatorname{diag}\left( \bm\Sigma^{\bm w,\text{DO}}, \bm\Sigma^{\bm w,\text{DO}} ,\dots \right)}$, yielding the distribution of the DO state sequence  as 
\begin{align}\label{eqn:covmat_XiDO}
     \bm\Xi^\text{DO}_N  \sim \mathcal{N}\left(\bar{\bm\Xi}^\text{DO}_N,   \bm B^\text{DO}   \bm \Sigma^{\bm W,\text{DO}}  {\bm B^\text{DO}}^\top  \right).
\end{align}
The offset $\bm q_N$ of \eqref{eqn:feasible_set} depends on the DO state sequence and therefore the covariance matrix for the offset $\bm \Sigma^{\bm{q}_N}$ is  
\begin{equation}\label{eqn:cvpmCovariancematrix}
    \bm\Sigma^{\bm{q}_N} =  \bm{Q}_N \bm B^\text{DO}   \bm \Sigma^{\bm W,\text{DO}}  {\bm B^\text{DO}}^\top  \bm{Q}_N^\top,
\end{equation}
and the mean of the offset vector $\bar{\bm{q}}_N$ is based on~\cite{brudigam2023}. 

With the stochastic properties of the offset vector, the constraint violation probability is 
\begin{IEEEeqnarray}{c}
    \Pr\big[\Xii_k\!\notin\!\hat{\mathcal{S}}(\hat{\bm\Xi}^\text{DO}_k)\big] 
    \!= \!1 - c\!
    \int_{\bm{S}_N\leq 0} \hspace{-1em}
    e^{- \tfrac{1}{2}\left\| \bm{S}_N- \bm{\mu}_k \right\|^2_{ {\bm \Sigma^{\bm{q}_{N}}}^{\text{-}1}}  }
    d\bm{S}_N,
    \label{eqn:constraint_satisfation_prob} \IEEEeqnarraynumspace
\end{IEEEeqnarray}
where $c= \frac{1}{\sqrt{(2\pi)^{4N}|\bm\Sigma^{\bm{q}_{N}}|}} $ and  $\bm{\mu}_k =\bm{Q}_N\Xii_N+\bar{\bm{q}}_N$. The constraint violation probability is determined by the integration of the constraint satisfaction probability \eqref{eqn:constraint_satisfation_prob}. The integration variable $\bm{S}_N\in\mathbb{R}^{4N}$ is non-positive to integrate over the space where the left side of the constraint ${\bm{Q}_N\Xii_N+\bm{q}_N}$ is negative, i.e., integration over the space where the constraint is fulfilled. 
We are not interested in the minimum value of~\eqref{eqn:cvpm_probablistic_case}, rather in the minimizer $\UU_N^*$. 
Therefore, the numerical integration in~\eqref{eqn:constraint_satisfation_prob} is approximated by a quadratic program in \eqref{eqn:approx_probabilistic_case} that yields an approximated minimizer \cite{fink2024}. 
 The approximation leads to a minimization of the probability by finding a mean value that is close to the boundary of the constraints based on a distance weighted by the covariance matrix. This method ensures that the solution is centered around the most probable scenarios for satisfying the constraints, rather than setting a strict upper or lower bound.
The solver uses the integration variable $\bm{S}_N$ as decision variable, finding the smallest weighted distance, which is the dominant part in the integral~\eqref{eqn:constraint_satisfation_prob}.

\section{Simulation results}
\label{sec:simulations}
We validate the proposed \smpcvpm~scheme through numerical simulations from the CommonRoad database~\cite{althoff2017}. 
We compare the performance of our novel method to that of methods considering a pre-stored fail-safe trajectory as emergency backup plan, i.e.,   fail-safe trajectory planning (FTP)~\cite{pek2018,brudigam2023}. Specifically, we tested the \smpcftp~scheme from~\cite{brudigam2023}, where, to facilitate the comparison, FTP is implemented using CVPM robust case.

The code is implemented in Python and both the set computations and the solution to all OCPs are computed using \textit{CVXPY}~\cite{diamond2016}. The EV trajectory is replanned every $T=0.1\ \si{s}$, since this is the sampling time of the trajectories in CommonRoad. Simulations were run on a computer with an AMD Ryzen 5 3500U eight-core processor.

We analyze the simulations of two challenging scenarios, selected because they consider trajectories of the DOs that trigger a reaction of the EV suitable to analyze the properties of our algorithm. 
We compare the performance of the control schemes by means of two metrics:
\begin{itemize}
    \item the computation time, which is of primary concern for the real-time applicability of the algorithms, since the input must be computed within short sampling times to iteratively re-plan the optimal trajectory of the EV;
    \item the optimality of the closed-loop trajectory, assessed by computing the average stage cost:
    \begin{equation}
    J_\text{sim} = \sum_{t=1}^{N_\text{sim}}\frac{\left(\|\Delta\xii(t)\|^2_{\bm{Q}}+\|\bm{u}(t)\|^2_{\bm{R}}\right)}{N_\text{sim}},
    \label{eqn:metrics_cost}%
\end{equation}%
where $N_\text{sim}$ is the number of steps in the scenario.
\end{itemize}

The values of the weighing matrices $\bm{Q}$, $\bm{R}$, and the reference target state $\xii^\ast$ in ${\Delta\xii\!=\!\xii\!-\!\xii^\ast}$, are as in the SMPC OCP~\eqref{eqn:costfcn}. Thus, $J_\text{sim}$ evaluates the optimality of the algorithms in closed loop as for the efficiency goals of the SMPC.

\subsection{Overtaking Maneuver} 
In the CommonRoad commonroad scenario \textit{DEU\_A99-1\_2\_T-1}, the EV is located on the right-most lane of the highway, surrounded by five other vehicles. The initial condition is represented in Figure~\ref{fig:DEUA99_iter0}. The vehicle located in the middle lane, to the left of the EV, performs an overtaking maneuver.

\begin{figure}
    \centering
    \vspace{-3mm}
    \subfloat[$t=\SI{0}{s}$ \label{fig:DEUA99_iter0}]{
        \hspace{-3mm}\resizebox{0.52\columnwidth}{!}{\input{pgf2/03-08-2023_10_40_04_DEU_A99-1_2_T-1_frame_1.pgf}}
    }
    \subfloat[$t=\SI{2.2}{s}$  \label{fig:DEUA99_iter23}]{
        \hspace{-4mm}\resizebox{0.52\columnwidth}{!}{\input{pgf2/03-08-2023_10_40_04_DEU_A99-1_2_T-1_frame_23.pgf}}
    }
    \caption{Traffic configuration of CommonRoad scenario \textit{DEU\_A99-1\_2\_T-1}. The EV is represented in red. The driving direction is from left to right. Faded colors show the positions of vehicles in the following two steps.}
\end{figure}
 
At the beginning, the proposed \smpcvpm~control scheme applies the SMPC control input, since the SMPC OCP is feasible and the safety check is successful, i.e., the SMPC solution is safe. Thus, the \smpcvpm~control scheme takes advantage of the optimistic planning of SMPC.
At simulation time $t=\SI{2.2}{s}$, see Figure~\ref{fig:DEUA99_iter23}, the vehicle performing the overtaking maneuver passes the EV and the optimal input obtained by SMPC fails in the safety check, that is, applying the SMPC solution might lead to a collision and the CVPM module is used.
In this case, there exists a solution with zero probability of constraint violation (robust case). In fact, the behavior of the overtaking vehicle did not violate the assumptions on the model. Hence, CVPM is used only to prevent the risk of collision that SMPC would lead to, and a solution with zero probability of collision is obtained.

In turn, at the beginning the \smpcftp~scheme~\cite{brudigam2023} also always finds the SMPC solution to be safe and applies it. Yet, this is only possible after that at each sampling time a safety backup is computed and stored. At the end, the SMPC solution is not guaranteed to be safe, thus the algorithm applies the backup stored during the previous iteration.

In this scenario in which no unanticipated situation happens our novel \smpcvpm~scheme yields an average stage cost~\eqref{eqn:metrics_cost} ${J_\text{sim}=0.283}$, which is marginally lower than the cost ${J_\text{sim}=0.302}$ of the \smpcftp~scheme~\cite{brudigam2023}. Further, we analyzed the average computation time for each module used, considering 100 simulations of the scenario. The average computation time for the SMPC branch for the \smpcftp~scheme~\cite{brudigam2023} is \SI{12.287}{\ms}, with \SI{3.158}{\ms} to solve the $\text{SMPC}(\bm x)$ problem and \SI{10.144}{\ms} to solve the $\text{FTP}(\bm x^+)$ problem. By contrast, in the \smpcvpm~scheme \SI{3.873}{\ms} are required for the SMPC branch, namely \SI{3.401}{\ms} for the $\text{SMPC}(\bm x)$ problem and just \SI{0.525}{\ms} for the $\text{Safety check}(\bm x^+)$, and \SI{0.554}{\ms} for the CVPM branch. Since the problems are solved in parallel, our scheme requires only \SI{3.873}{\ms}, with a reduction of 70\% with respect to the \smpcftp~scheme~\cite{brudigam2023}.
The increased efficiency lies in the fact that the \smpcftp~scheme~\cite{brudigam2023}, similar to~\cite{pek2018}, requires the solution of two cascaded OCPs ($\text{SMPC}(\bm x)$ and $\text{FTP}(\bm x^+)$), whereas in our \smpcvpm~scheme $\text{SMPC}(\bm x)$ is followed by a quick set computation ($\text{Safety check}(\bm x^+)$). 
Although it requires less computation time, our scheme still only applies the SMPC solution after ensuring it is safe.

\subsection{Critical Emergency Braking on a Multi-Lane Highway}
\begin{figure}
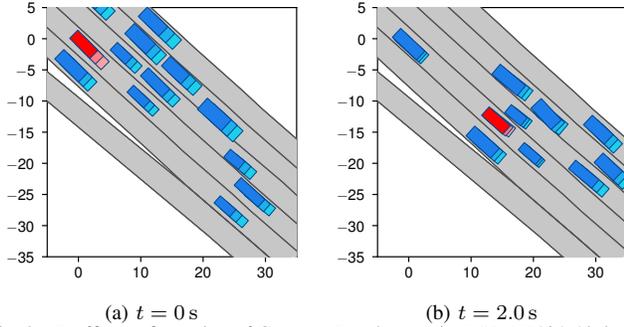

    \centering
    \subfloat[$t=\SI{0}{s}$\label{fig:USAUS101_iter0}]{
        \resizebox{0.48\columnwidth}{!}{\input{pgf2/03-08-2023_10_41_52_USA_US101-13_2_T-1_frame_1.pgf}}
    }
    \subfloat[$t=\SI{2.0}{s}$\label{fig:USAUS101_iter20}]{
        \resizebox{0.48\columnwidth}{!}{\input{pgf2/03-08-2023_10_41_52_USA_US101-13_2_T-1_frame_20.pgf}}
    }
    \caption{Traffic configuration of CommonRoad scenario \textit{USA\_US101-13\_2\_T-1}. The EV is represented in red. The driving direction is from top-left to bottom-right. Faded colors show the positions of vehicles in the following two steps.}
\end{figure}

In the CommonRoad simulation \textit{USA\_US101-13\_2\_T-1}, the EV is located on the second lane from the right of the highway, surrounded by thirteen other vehicle, see Figure~\ref{fig:USAUS101_iter0}. In the scenario, the traffic comes to a critical stop, forcing the EV to perform an emergency break in order to minimize the probability of collision with surrounding vehicles.

\begin{figure}
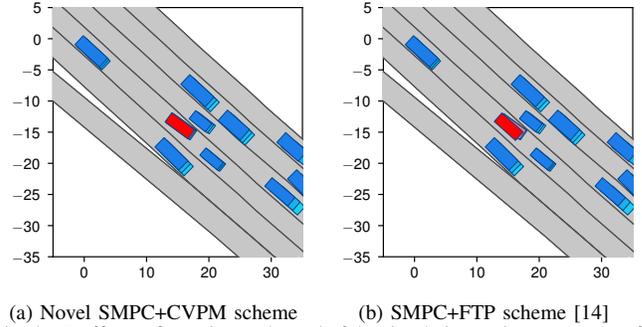

    \centering
    \subfloat[Novel \smpcvpm~scheme  \label{fig:USAUS101_iter27}]{
        \resizebox{0.48\columnwidth}{!}{\input{pgf2/03-08-2023_10_41_52_USA_US101-13_2_T-1_frame_24.pgf}}
    }
    \subfloat[\smpcftp~scheme~\cite{brudigam2023}  \label{fig:USAUS101_iter27_tim}]{
        \resizebox{0.48\columnwidth}{!}{\input{pgf2/03-08-2023_10_41_52_USA_US101-13_2_T-1_frame_24_tim.pgf}}
    }
    \caption{Traffic configuration at the end of the simulation at time $t=\SI{2.4}{s}$ of CommonRoad scenario \textit{USA\_US101-13\_2\_T-1} used in the second simulation.
    }
\end{figure}

During the initial part of the braking maneuver, our novel \smpcvpm~scheme applies the SMPC solution, as the safety check succeeds. Thus, at first the algorithm benefits from the efficient planning of SMPC. Yet, from simulation time $t=\SI{2.0}{s}$, represented in Figure~\ref{fig:USAUS101_iter20}, the SMPC OCP continues to provide a solution but the safety check fails, i.e., the SMPC solution might result in collisions.
Hence, the solution provided by the CVPM branch is applied, obtained in parallel to the SMPC branch. Here, no solution with zero probability of constraint violation exists, because the breaking maneuver of the DOs is very sudden and the actual trajectory violates the assumptions on the DO behavior from Sec.~\ref{sec:dynamic_models}. Thus, the probabilistic case is used, i.e., the \smpcvpm~scheme determines a new sequence of input explicitly aiming at minimizing the overall probability of collision. In this scenario, the trajectory consists of a braking and steering maneuver to the left, similar to the decision of several other human drivers in the recorded scenario. The position and heading of the EV at the end of the simulation is shown in Figure~\ref{fig:USAUS101_iter27}.

At first, the \smpcftp~scheme~\cite{brudigam2023} also applies the SMPC solution, after computing a safety backup. 
When the SMPC solution is not guaranteed to be safe, the \smpcftp~scheme~\cite{brudigam2023} relies on the safety backup stored during the previous iterations, consists of a rapid deceleration while remaining at the center of the lane. Figure~\ref{fig:USAUS101_iter27_tim} shows the position and heading of the EV at the end of the simulation. Differently from the solution of our \smpcvpm~scheme (Figure~\ref{fig:USAUS101_iter27}), here the EV does not turn to the left to minimize the probability of collisions with DOs.

In this scenario the \smpcftp~scheme~\cite{brudigam2023} results in an average stage cost~\eqref{eqn:metrics_cost} of $J_\text{sim}=102.9$, whereas the novel \smpcvpm~scheme yields a marginally lower cost, $J_\text{sim}=100.7$. 
Moreover, we analyzed again the average computation time for each module used, considering 100 simulations of the scenario. For the \smpcftp~scheme~\cite{brudigam2023}, SMPC branch is solved in \SI{23.115}{\ms} on the average, with \SI{10.326}{\ms} to solve the $\text{SMPC}(\bm x)$ problem and \SI{12.789}{\ms} to solve the $\text{FTP}(\bm x^+)$ problem. In the \smpcvpm~scheme, just \SI{10.181}{\ms} are required for the SMPC branch, namely \SI{9.434}{\ms} for the $\text{SMPC}(\bm x)$ problem and \SI{0.746}{\ms} for the $\text{Safety check}(\bm x^+)$. The CVPM branch is solved in \SI{2.068}{\ms} on the average, with \SI{0.753}{\ms} in Check case $\text{CVPM}(\bm x)$ and \SI{1.316}{\ms} in $\text{CVPM Prob}(\bm x)$. Since the problems are solved in parallel, our scheme requires only \SI{10.181}{\ms}, with a reduction of 55\% with respect to the \smpcftp~scheme~\cite{brudigam2023}. 
As computation time scales significantly with scenario complexity, such as a larger number of surrounding vehicles, this reduction can be crucial for practical application.

\section{Discussion}
\label{sec:discussion}
One major advantage of the novel \smpcvpm~scheme is that unanticipated situations, for which a feasible and safe solution cannot be determined, are explicitly addressed with a view at minimizing the probability of collision. Though intuitively reasonable, minimizing the probability of collision should not be the only objective of the EV. Between a first trajectory yielding a $40\%$ probability of collision with anticipated $10\%$ severity of the impact and a second trajectory in which the probability of collision is $30\%$ and the expected severity of the impact is $20\%$, it is not obvious that the latter trajectory is preferable. The anticipated harm caused by a collision is to be taken into account. Sometimes choosing the trajectory yielding a higher probability of collision proves beneficial, if this in turns significantly reduces the severity caused by the collision. To balance among different types of conflicting objectives, i.e., minimizing the probability of collision and the anticipated severity of the impact, respectively, an ad-hoc cost function could be designed and used in the CVPM probabilistic case. The balance between these objectives is worth a special discussion and is not a purely technical matter, rather ethical, related to the well-known Trolley Problem~\cite{geisslinger2021}. 

Compared to schemes relying on a pre-stored fail-safe trajectory~\cite{pek2018,brudigam2023}, our novel \smpcvpm~scheme offers several advantages. Concerning the computation complexity, the \smpcftp~scheme~\cite{brudigam2023} intrinsically requires the cascaded solution of two OCPs, since to guarantee safety a backup for the next iteration must be stored. For short sampling times, the solution of two cascaded optimization problems is challenging. The scheme from~\cite{alsterda2019} computes jointly the nominal and the backup trajectory, significantly increasing the size of the OCP and, thus, the computation time. In turn, our novel scheme does not require the computation of a backup and requires a rather smaller computation time. In practice, it is not necessary to pre-compute the backup, rather it suffices to verify that from the first predicted state of the SMPC trajectory a safe solution can be obtained. If necessary, such safe solution will be quickly computed at the next iteration by the robust case of CVPM, in parallel to the SMPC branch. Hence, the increased efficiency of our scheme does not compromise safety.

Besides, in presence of unanticipated behaviors of DOs that prevents the existence of a robust safe trajectory, applying the backup derived at the previous iteration~\cite{brudigam2023} is not safe, as the safety backup is a valid and safe trajectory only under the assumption that the DOs behave as expected at the previous iteration. Hence, the scheme does not react to the present traffic configuration, which is hazardous and could lead to danger in practice if DOs exhibit unanticipated behaviors. At best, this leads to suboptimal behaviors in terms of minimization of the probability of collisions. In turn, our \smpcvpm~scheme uses the probabilistic case of CVPM to plan the trajectory minimizing the probability of constraint violation.

\section{Conclusion}
\label{sec:conclusion}
A novel scheme for safe trajectory planning is introduced, which permits to benefit from the efficient planning of SMPC. However, the sequence of inputs obtained by the SMPC planner is applied only if safe, i.e., if it will lead to a state from which it is possible to compute a feasible trajectory robust against the uncertainty. Therefore, the EV does not cause a collision for anticipated behaviors of traffic participants. In the event of unanticipated behaviors of traffic participants, the scheme yields a trajectory obtained by minimizing the probability of collision. The scheme is highly efficient, as the two OCPs can be solved in parallel and the approach can be modified to account for the harm of the collision. Numerical simulations based on scenarios drawn from the benchmark dataset CommonRoad~\cite{althoff2017} and comparing with the performance of the previous scheme from~\cite{brudigam2023} demonstrated the benefit of our approach. The presented \smpcvpm~scheme is suitable for application beyond autonomous driving, especially in frameworks characterized by large uncertainty because of the interaction with humans, e.g., human-robot-collaboration.

\section*{Acknowledgments}
The authors thank Tim Br\"{u}digam for valuable discussions.

\bibliographystyle{IEEEtran}
\bibliography{res}

\end{document}

%% file: space.tex
\belowdisplayskip \abovedisplayskip%
\skip\footins 0.5\baselineskip  plus 0.4\baselineskip  minus 0.2\baselineskip

%% file: Figs/Diagram_logic.tex
\begin{tikzpicture}

\def \nsh {0.8}         
\def \nsw {1.5}         
\def \smpch {1.0}       
\def \smpcw {1.8}       
\def \safetyh {1.0}     
\def \safetyw {1.9}     
\def \cvpmh {1.0}       
\def \cvpmw {2.5}       
\def \caseAh {1.0}      
\def \caseAw {2.5}      
\def \caseBh {1.0}      
\def \caseBw {2.2}      
\def \applysmpch {0.8}  
\def \applysmpcw {2.1}  
\def \applycvpmh {0.8}  
\def \applycvpmw {2.1}  
\def \dh {1.7}          
\def \dhh {1.3}          
\def \dw {2.8}            
\def \ddw {1.3}         
\def \roundness {0.35cm}
\def \colSMPC {blue}
\def \colCVPM {red}
\def \nst {\small New state}                                
\def \nstosmpc {$\bm x$}                                        
\def \smpct {\small Solve\\\small $\text{SMPC}(\bm x)$}         
\def \smpctosafety {\small feasible, $\bm x^+$}                 
\def \safetyt {\small Safety\\\small check $\bm x^+$}           
\def \safetytoapplysmpc {\small safe}                       
\def \safetytocvpm {\small unsafe}                          
\def \smpctocvpm {\small infeasible}                        
\def \cvpmt {\small Is robust\\\small planner feasible?}    
\def \caseAt {\small Solve\\\small $\text{CVPM Robust}(\bm x)$}                  
\def \caseBt {\small Solve\\\small $\text{CVPM Prob}(x)$}                  
\def \applysmpct {\small Apply SMPC}                        
\def \applycvpmt {\small Apply CVPM}                        

\draw[rounded corners=\roundness,thick] (-\nsw/2,-\nsw/2) rectangle ++(\nsw,\nsh) node[pos=0.5,align=center] {\nst};
\draw[->,thick] (0,-\nsw/2) to node[pos=0.5,fill=white, anchor=west] {\nstosmpc} (0,-\dh+\smpch/2);
\draw[rounded corners=\roundness,fill=\colSMPC!10,draw=\colSMPC,thick] (-\smpcw/2,-\smpch/2-\dh) rectangle ++(\smpcw,\smpch) node[pos=0.5,align=center] {\smpct};
\draw[->,thick] (0,-\dh-\smpch/2) to node[pos=0.5,fill=white, anchor=south east] {\smpctosafety} (-\dw,-2*\dh+\safetyh/2);
\draw[rounded corners=\roundness,fill=\colSMPC!10,draw=\colSMPC,thick] (-\safetyw/2-\dw,-\safetyh/2-2*\dh) rectangle ++(\safetyw,\safetyh) node[pos=0.5,align=center] {\safetyt};
\draw[->,thick] (-\dw,-2*\dh-\safetyh/2) to node[pos=0.5,fill=white] {\safetytoapplysmpc} (-\dw,-2*\dh-2*\dhh+\applysmpch/2);
\draw[rounded corners=\roundness,fill=\colSMPC!10,draw=\colSMPC,thick] (-\applysmpcw/2-\dw,-\applysmpch/2-2*\dh-2*\dhh) rectangle ++(\applysmpcw,\applysmpch) node[pos=0.5,align=center] {\applysmpct};

\draw[->,thick] (0,-\dh-\smpch/2) to node[pos=0.5,fill=white, anchor=south west] {\smpctocvpm} (\dw,-2*\dh+\cvpmh/2);
\draw[->,thick] (-\dw+\safetyw/2,-2*\dh) to node[pos=0.5,fill=white] {\safetytocvpm} (\dw-\cvpmw/2,-2*\dh);
\draw[rounded corners=\roundness,fill=\colCVPM!10,draw=\colCVPM,thick] (-\cvpmw/2+\dw,-\cvpmh/2-2*\dh) rectangle ++(\cvpmw,\cvpmh) node[pos=0.5,align=center] {\cvpmt};
\draw[->,thick] (\dw,-2*\dh-\cvpmh/2) to (\dw-\ddw,-2*\dh-1*\dhh+\caseAh/2);
\draw[rounded corners=\roundness,fill=\colCVPM!10,draw=\colCVPM,thick] (-\caseAw/2+\dw-\ddw,-\caseAh/2-2*\dh-1*\dhh) rectangle ++(\caseAw,\caseAh) node[pos=0.5,align=center] {\caseAt};
\draw[->,thick] (\dw-\ddw,-2*\dh-1*\dhh-\caseAh/2) to (\dw-\ddw/3,-2*\dh-2*\dhh+\applycvpmh/2);
\draw[->,thick] (\dw,-2*\dh-\caseBh/2) to (\dw+\ddw,-2*\dh-1*\dhh+\caseBh/2);
\draw[rounded corners=\roundness,fill=\colCVPM!10,draw=\colCVPM,thick] (-\caseBw/2+\dw+\ddw,-\caseBh/2-2*\dh-1*\dhh) rectangle ++(\caseBw,\caseBh) node[pos=0.5,align=center] {\caseBt};
\draw[->,thick] (\dw+\ddw,-2*\dh-1*\dhh-\caseBh/2) to (\dw+\ddw/3,-2*\dh-2*\dhh+\applycvpmh/2);
\draw[rounded corners=\roundness,fill=\colCVPM!10,draw=\colCVPM,thick] (-\applycvpmw/2+\dw,-\applycvpmh/2-2*\dh-2*\dhh) rectangle ++(\applycvpmw,\applycvpmh) node[pos=0.5,align=center] {\applycvpmt};

\end{tikzpicture}
\vspace{-12mm}

%% file: output.bbl
\begin{thebibliography}{10}
\providecommand{\url}[1]{#1}
\csname url@samestyle\endcsname
\providecommand{\newblock}{\relax}
\providecommand{\bibinfo}[2]{#2}
\providecommand{\BIBentrySTDinterwordspacing}{\spaceskip=0pt\relax}
\providecommand{\BIBentryALTinterwordstretchfactor}{4}
\providecommand{\BIBentryALTinterwordspacing}{\spaceskip=\fontdimen2\font plus
\BIBentryALTinterwordstretchfactor\fontdimen3\font minus \fontdimen4\font\relax}
\providecommand{\BIBforeignlanguage}[2]{{%
\expandafter\ifx\csname l@#1\endcsname\relax
\typeout{** WARNING: IEEEtran.bst: No hyphenation pattern has been}%
\typeout{** loaded for the language `#1'. Using the pattern for}%
\typeout{** the default language instead.}%
\else
\language=\csname l@#1\endcsname
\fi
#2}}
\providecommand{\BIBdecl}{\relax}
\BIBdecl

\bibitem{kumar2013}
P.~Kumar, M.~Perrollaz, S.~Lef{\`e}vre, and C.~Laugier, ``Learning-based approach for online lane change intention prediction,'' in \emph{{{IEEE Intelligent Vehicles Symposium}} {{IV}}}, 2013.

\bibitem{phillips2017}
D.~Phillips, T.~Wheeler, and M.~Kochenderfer, ``Generalizable intention prediction of human drivers at intersections,'' in \emph{{{IEEE}} {{IV}}}, 2017.

\bibitem{rosolia2017}
U.~Rosolia, A.~Carvalho, and F.~Borrelli, ``Autonomous racing using learning {{Model Predictive Control}},'' in \emph{{{ACC}}}, 2017.

\bibitem{levinson2011}
J.~Levinson, J.~Askeland, J.~Becker, J.~Dolson, D.~Held, S.~Kammel, J.~Z. Kolter, D.~Langer, O.~Pink, V.~Pratt, M.~Sokolsky, G.~Stanek, D.~Stavens, A.~Teichman, M.~Werling, and S.~Thrun, ``Towards fully autonomous driving: {{Systems}} and algorithms,'' in \emph{{{IEEE}} {{IV}}}, 2011.

\bibitem{carvalho2014}
A.~Carvalho, Y.~Gao, S.~Lefevre, and F.~Borrelli, ``Stochastic predictive control of autonomous vehicles in uncertain environments,'' in \emph{International {{Symposium}} on {{Advanced Vehicle Control}}}, 2014.

\bibitem{gutjahr2017}
B.~Gutjahr, L.~Gr{\"o}ll, and M.~Werling, ``Lateral {{Vehicle Trajectory Optimization Using Constrained Linear Time-Varying MPC}},'' \emph{IEEE Transactions on Intelligent Transportation Systems}, 2017.

\bibitem{dixit2020}
S.~Dixit, U.~Montanaro, M.~Dianati, D.~Oxtoby, T.~Mizutani, A.~Mouzakitis, and S.~Fallah, ``Trajectory {{Planning}} for {{Autonomous High-Speed Overtaking}} in {{Structured Environments Using Robust MPC}},'' \emph{IEEE Transactions on Intelligent Transportation Systems}, 2020.

\bibitem{soloperto2019}
R.~Soloperto, J.~K{\"o}hler, F.~Allg{\"o}wer, and M.~M{\"u}ller, ``Collision avoidance for uncertain nonlinear systems with moving obstacles using robust {{Model Predictive Control}},'' in \emph{{{ECC}}}, 2019.

\bibitem{sontges2018}
S.~S{\"o}ntges and M.~Althoff, ``Computing the {{Drivable Area}} of {{Autonomous Road Vehicles}} in {{Dynamic Road Scenes}},'' \emph{IEEE T-ITS}, 2018.

\bibitem{gillula2013}
J.~H. Gillula, C.~J. Tomlin, R.~P. Fedkiw, and J.-C. Latombe, ``Guaranteeing safe online machine learning via reachability analysis,'' Ph.D. dissertation, Stanford University, 2013.

\bibitem{schurmann2018}
B.~Sch{\"u}rmann, N.~Kochdumper, and M.~Althoff, ``Reachset {{Model Predictive Control}} for {{Disturbed Nonlinear Systems}},'' in \emph{2018 {{IEEE Conference}} on {{Decision}} and {{Control}} ({{CDC}})}, 2018.

\bibitem{mesbah2016}
A.~Mesbah, ``Stochastic {{Model Predictive Control}}: {{An Overview}} and {{Perspectives}} for {{Future Research}},'' \emph{IEEE Control Systems Magazine}, 2016.

\bibitem{farina2016}
M.~Farina, L.~Giulioni, and R.~Scattolini, ``Stochastic linear {{Model Predictive Control}} with chance constraints -- {{A}} review,'' \emph{Journal of Process Control}, 2016.

\bibitem{brudigam2023}
T.~Br{\"u}digam, M.~Olbrich, D.~Wollherr, and M.~Leibold, ``Stochastic {{Model Predictive Control}} with a {{Safety Guarantee}} for {{Automated Driving}},'' \emph{IEEE Transactions on Intelligent Vehicles}, 2023.

\bibitem{alsterda2019}
J.~P. Alsterda, M.~Brown, and J.~C. Gerdes, ``Contingency {{Model Predictive Control}} for {{Automated Vehicles}},'' in \emph{{{ACC}}}, 2019.

\bibitem{brudigam2021c}
T.~Br{\"u}digam, V.~Ga{\ss}mann, D.~Wollherr, and M.~Leibold, ``Minimization of constraint violation probability in model predictive control,'' \emph{International Journal of Robust and Nonlinear Control}, 2021.

\bibitem{fink2022}
M.~Fink, T.~Br{\"u}digam, D.~Wollherr, and M.~Leibold, ``Constraint {{Violation Probability Minimization}} for {{Norm-Constrained Linear Model Predictive Control}},'' in \emph{2022 {{European Control Conference}} ({{ECC}})}, 2022.

\bibitem{althoff2017}
M.~Althoff, M.~Koschi, and S.~Manzinger, ``{{CommonRoad}}: {{Composable}} benchmarks for motion planning on roads,'' in \emph{{{IEEE}} {{IV}}}, 2017.

\bibitem{werling2008}
M.~Werling and L.~Groll, ``Low-level controllers realizing high-level decisions in an autonomous vehicle,'' in \emph{{{IEEE}} {{IV}}}, 2008.

\bibitem{benciolini2023}
T.~Benciolini, D.~Wollherr, and M.~Leibold, ``Non-{{Conservative Trajectory Planning}} for {{Automated Vehicles}} by {{Estimating Intentions}} of {{Dynamic Obstacles}},'' \emph{IEEE Transactions on Intelligent Vehicles}, 2023.

\bibitem{borrelli2017}
F.~Borrelli, A.~Bemporad, and M.~Morari, \emph{Predictive {{Control}} for {{Linear}} and {{Hybrid Systems}}}.\hskip 1em plus 0.5em minus 0.4em\relax Cambridge University Press, 2017.

\bibitem{fink2024}
\BIBentryALTinterwordspacing
M.~Fink, T.~Br{\"u}digam, D.~Wollherr, and M.~Leibold, ``Minimal {{Constraint Violation Probability}} in {{Model Predictive Control}} for {{Linear Systems}},'' 2024, accepted at IEEE-TAC. [Online]. Available: \url{http://arxiv.org/abs/2402.10538}
\BIBentrySTDinterwordspacing

\bibitem{pek2018}
C.~Pek and M.~Althoff, ``Computationally {{Efficient Fail-safe Trajectory Planning}} for {{Self-driving Vehicles Using Convex Optimization}},'' in \emph{{{International Conference}} on {{Intelligent Transportation Systems}}}, 2018.

\bibitem{diamond2016}
S.~Diamond and S.~Boyd, ``{{CVXPY}}: {{A Python-Embedded Modeling Language}} for {{Convex Optimization}},'' \emph{Journal of Machine Learning Research}, 2016.

\bibitem{geisslinger2021}
M.~Geisslinger, F.~Poszler, J.~Betz, C.~L{\"u}tge, and M.~Lienkamp, ``Autonomous {{Driving Ethics}}: From {{Trolley Problem}} to {{Ethics}} of {{Risk}},'' \emph{Philosophy \& Technology}, 2021.

\end{thebibliography}
